# Spin-Exchange Interaction in ZnO-based Quantum Wells

B. Gil, P. Lefebvre, T. Bretagnon, T. Guillet, J.A. Sans*, and T. Taliercio

*Université Montpellier 2, Place Eugène Bataillon, Groupe d'Etude des Semiconducteurs, case courrier 074, F-34095 Montpellier cedex 5, France*

C.Morhain

*Centre de Recherche sur l'Hétéro-Epitaxie et Applications, Rue Bernard Gregory, Sophia Antipolis, F-06560 Valbonne, France*

Abstract: Wurtzitic ZnO/(Zn,Mg)O quantum wells grown along the (0001) direction permit unprecedented tunability of the short-range spin exchange interaction. In the context of large exciton binding energies and electron-hole exchange interaction in ZnO, this tunability results from the competition between quantum confinement and giant quantum confined Stark effect. By using time-resolved photoluminescence we identify, for well widths under 3 nm, the redistribution of oscillator strengths between the A and B excitonic transitions, due to the enhancement of the exchange interaction. Conversely, for wider wells, the redistribution is cancelled by the dominant effect of internal electric fields, which dramatically reduces the exchange energy.

* Permanent address : *Institut de Ciència dels Materials, Departament de Física Aplicada, Universitat de València, Ed. Investigació, E-46100 Burjassot, Spain*

Exchange interaction energy naturally appears in quantum mechanics whenever one has to deal with identical interacting particles. This interaction is important for understanding particle physics, interpreting small energy splittings in atoms and for describing spin-related excitonic fine structure in condensed matter. In this latter case, a prerequisite to the interpretation of the overall optical properties of pure solids has been the definition of the short-range and long-range contributions to this interaction [1].

In this paper we address the first report on a modification of the short range electron-hole exchange interaction with well width, in ZnO-ZnMgO wurtzitic quantum wells (WQWs). We provide a good description of our results by using a three-band model that is an extension of the model found to be useful for a quantitative analysis of the optical properties of bulk ZnO [2, 3]. Due to an efficient Quantum Confined Stark Effect (QCSE), the short-range electron-hole exchange interaction redistributes oscillator strengths among exciton states built from different valence bands in a way that depends on well width.

For group III-nitride WQWs, due to excessive inhomogeneous broadening effects [4], the optical properties and their evolution with time are treated in the context of a band to band model. The short-range electron-hole exchange interaction is not identified as an important issue although it plays, for micrometric epilayers [5], a role in the mixing of the excitonic states that originate from the different valence-bands. The short-range electron-hole exchange interaction $\gamma$ equals 0.69 meV for GaN [5] and 0.02 meV for GaAs [6]. Gil [3] has established for ZnO, by using a model including the appropriate excitonic symmetry, *that the weak oscillator strength of the low energy A-exciton compared to the oscillator strength of the higher energy B-exciton results, in particular, from quantum beat oscillations that are extremely efficient due to the giant magnitude of the short-range electron-hole exchange interaction energy* ($\gamma = 4.7$ meV in ZnO). Such quantum beat oscillations cannot be studied in the context of a band to band model. This indicates that the ZnO-ZnMgO WQW is the adequate system for proving the need to use *a true excitonic description,* since huge modifications of the optical properties cannot be interpreted out of the context of a model that includes the short-range exchange interaction *and* the tuning of its magnitude by varying well widths.

When reducing the width of a quantum well ($L_W$), for a recombination process, one generally observes a blue-shift of the emission, as well as an enhancement of the excitonic binding energy [7], and of the electron-hole exchange interaction [8, 9]. However, internal electric fields which are spontaneously generated in (0001)-grown WQW structures, from the breaking of on-axis translational symmetry at the hetero-interfaces [10] produce a huge QCSE that drastically reduces the excitonic oscillator strength by separating the wave-functions of the electron and hole while smoothly reducing the excitonic binding energy via the extension of the in-plane relative motion of the electron-hole pair [11, 12].

Let us first consider group-III nitride semiconductors for comparison. They exhibit particularly large coefficients of both spontaneous and piezoelectric polarizations (comparable to those of ZnO) [13-15] and thus can generate internal fields of several MV/cm. For wide enough nitride WQWs (above a few nanometers, in practice), the QCSE is so strong that it can counteract the quantum confinement and therefore can push the fundamental transition *below* the natural excitonic gap of the WQW material [14]. Two regimes of well widths can be defined:

- for narrow WQWs, the quantum confinement dominates, yielding large emission energies and typical excitonic lifetimes of nanoseconds, or slightly less,

- for wide enough WQWs, the QCSE dominates and the emission energy decreases almost linearly with increasing $L_W$, whereas the exciton lifetime increases (its oscillator strength decreases) almost exponentially.

Similar effects are expected in case of ZnO / (Zn,Mg)O, but enriched from short-range interaction effects.

Quantum confinement is known to change this interaction by a factor that is inversely proportional to the squeezed volume occupied by the confined exciton wave-function. For quantum wells, the modification factor is given by [8]:

$$\alpha = \pi a_{3D}^{3} W_{eh} |\varphi_\lambda(0)|^2 \qquad (1),$$

where $a_{3D}$ is the excitonic effective Bohr radius for the bulk material. The exciton relative motion is described by the product of envelope-functions $f_e(z_e)$ and $f_h(z_h)$ describing the electron and hole confinement along the QW growth axis, by the two-dimensional hydrogenic term $\varphi_\lambda(\rho) = \left(\sqrt{2/\pi}/\lambda\right) \exp(-\rho/\lambda)$, where $\rho$ is the in-plane relative coordinate. The z-dependent factor is given by:

$$W_{eh} = \int_{-\infty}^{+\infty} dz \left[ f_e(z) f_h(z) \right]^2 \qquad (2).$$

From Eqs. (1) and (2) it is obvious that, in any case of on-axis separation of the electron and hole, the $\alpha.\gamma$ factor may become smaller than unity and even vanish with increasing $L_W$, in presence of a longitudinal electric field.

Due to large spontaneous and piezoelectric polarization parameters [15], WQWs based on ZnO, like ZnO-(Zn,Mg)O WQWs, are thus expected to materialize the unprecedented situation where *both* large internal electric fields *and* giant exchange interaction terms are present. To face this situation, competitive effects of confinement and electric fields on the large exciton binding energy (~60 meV), have to be properly considered.

The samples were grown by plasma-assisted Molecular beam Epitaxy, the metals (Zn and Mg) being evaporated using Knudsen cells and atomic O being activated in a radio-frequency plasma cell. After thermal cleaning of the c-oriented sapphire substrates, the latter were exposed to the O beam for 5-10 minutes prior the deposition of ZnO. The growth of a ZnO template was then carried out at a growth temperature of 520°C and a growth rate of ~ 0.45 µm/h. In order to minimize the effect of Al diffusion, which acts as a shallow donor, from the substrate and to ensure a low carrier concentration in the QW area, templates with thicknesses of 1µm were grown. Based on previous electrochemical capacitance-voltage measurements of ZnO epilayers grown using the same growth conditions, we estimate the residual carrier concentration in the vicinity of the QWs to be less than $10^{16}$ cm$^{-3}$. The QW heterostructures consist in the growth of a 0.2 µm-thick (Zn, Mg)O layer, a ZnO QW with thicknesses varying from sample to sample, and a 0.1 µm-thick (Zn, Mg)O cap layer. QW thicknesses were varied from 1.6 to 9.5 nm; the Mg content of the barrier layers was in the range of 21-22%, varying slightly from sample to sample. More details can be found in ref.16.

In the following, we use results of time-integrated PL, obtained under pulsed excitation, instead of cw PL. The 2-ps laser pulses were provided by the third harmonic (260 nm) of a titanium-sapphire laser, with typical energy densities of ~50 nJ/cm$^2$ per pulse. It is important to outline that unintended photo-induced screening of electric field easily occurs in such wide bandgap WQWS [13-15] and that correct observation of an unscreened situation (and correct measurement of a photoluminescence energy) requires to work using pulsed excitation conditions at a repetition rate slow enough to reach a situation where both exponential intensity decay time **and** absence of any red-shift of the photoluminescence are reached. Therefore the repetition rate was adapted to the slowness of the observed decay, by use of an acousto-optical modulator, between 800 Hz and 80 kHz, thus allowing *for a full de-excitation between two consecutive pulses*.

Figure 1 shows, for four different QW widths, the time-integrated photoluminescence (PL) spectra obtained under pulsed excitation. Inserts in Fig.1 show the band profiles, energy levels and envelope functions for electrons and heavy holes that we calculate for two extreme situations, namely narrow and wide QWs. A very good agreement is obtained between our variational "one band" excitonic calculation [15] and the experimental PL energies, as shown in Fig. 3 of reference 15, when we include an electric field of 0.9 MV/cm.

The well width dependence of the oscillator strength is usually considered to be the same as that of quantity $f = |\varphi_\lambda(0)|^2 I_{eh}^2$, where $I_{eh} = \int_{-\infty}^{+\infty} dz \, f_e(z) \, f_h(z)$ is the envelope-function overlap integral. So, in this first approach, we can readily see how fast the oscillator strength is expected to decrease: when the QCSE becomes dominant, we obtain a drop of more than 7 orders of magnitude between $L_W$ = 2 and 7 nm. Wide WQWs show very large PL decay times, at T = 8K, and they grow almost exponentially with $L_W$. As an illustration the data for the 7.1 nm wide WQW is shown in the right-hand plot in Fig. 2.

Experimentally, things are a little more complicated, especially for narrow WQWs, as reported in Ref. 15. The left hand plot in Fig. 2 indicates that the PL dynamics of WQWs emitting at energies above the bandgap of ZnO exhibit two characteristic times. Typically, a first decay occurs with a slightly sub-nanosecond time ($\tau_1$), followed by a second one, with decay time of a few tens of nanoseconds ($\tau_2$). We have checked that the slower part of this dynamics was not induced by a slow carrier transfer from the barriers: TR-PL from these barriers revealed much slower decay times, on orders of hundreds of microseconds. These $\tau_1$ and $\tau_2$ are accurately measured if recording the decays using different, appropriate, times scales as shown here in the case of the 1.6 nm thin WQW. The measured decay times are reported in Fig. 3 (dots).

We explain these results by the superposition of two radiative mechanisms related to the two fundamental excitonic ground states, usually referred to as the A and B excitons, in the bulk material. As explained above, the e-h exchange interaction mixes the $\Gamma_5$ exciton states, defined in Ref. 3, that are dipole allowed in the present σ-polarization [2, 3]. The short range exchange energy, $\gamma$ = 4.73 meV for bulk ZnO, strongly depends on well-width for the present WQWs, as illustrated by the right hand bottom insert in Fig. 3. which presents the changes of the e-h exchange energy, $\alpha.\gamma$ calculated from Eqs. (1) and (2), versus $L_W$. For narrow enough QWs, the confinement enhances both direct Coulomb (see reference 15) and short-range exchange energies. But they both become smaller than their bulk value for QW widths larger than ~2-3 nm.

For narrow QWs, the mixing of the two fundamental $\Gamma_5$ states by $\alpha.\gamma$ necessarily redistributes some oscillator strength from the A exciton to the benefit of the B exciton, like it is the case [3] for bulk ZnO, but more strongly, since in that case, the exchange energy $\alpha.\gamma$ is enhanced. To describe the case of our samples, we have used an adapted version of the formalism presented in Ref.3, for $\Gamma_5$ states. More specifically, as we did in Ref.15, we assume, for simplicity, equal on-axis effective masses for the so-called heavy- and light-holes. *This warrants that, even for QWs, the energy difference between diagonal terms for these two types of holes is independent of the well width.* We assume, too, that the bulk binding energies are equal for A, B and C excitons, for all QW widths. Again, this permits that the energy separation of diagonal terms in the $\Gamma_5$ excitonic Hamiltonian of Ref. 3 is kept constant. We assume, moreover, that internal strain effects are negligible, in first approximation, or that we can mimic them by our choice of the $\Delta_1$ crystal-field parameter. Finally, we consider a single confined state for electrons as well as for each of the three hole bands, which is reasonable in view of the band profiles sketched in Fig. 1.

With such hypotheses, we are left with the resolution of the following matrix:

$$\begin{matrix} |\Gamma_{5A}\rangle & |\Gamma_{5B}\rangle & |\Gamma_{5C}\rangle \end{matrix}$$
$$\begin{pmatrix} E^* - \Delta_1 - \Delta_2 + \alpha\gamma/2 & \alpha\gamma & 0 \\ \alpha\gamma & E^* - \Delta_1 + \Delta_2 + \alpha\gamma/2 & -\sqrt{2}\Delta_3 \\ 0 & -\sqrt{2}\Delta_3 & E^* - \alpha\gamma/2 \end{pmatrix} \quad (3).$$

Here, the bulk-like $\alpha\gamma$ term is simply the one that is plotted in the right hand bottom insert in Fig. 3 and $E^*$ is an energy which accounts for the band-gap, augmented by the confinement energies of the electron and the hole and reduced by the size-dependent binding energy. Within our approximations, the B line remains a few meV above the fundamental A line. In fact, some $L_W$ dependency (left hand top insert in Fig. 3) is only obtained for $L_W < 3$ nm and the maximum separation of these lines is of 12 meV, which is rather small, if compared to the PL linewidths for the two narrower QWs, comprised between 25 and 35 meV. These inhomogeneous linewidths result from the fluctuations of Mg concentration in the barriers and of strain along the QW plane. We thus understand that, for all well widths, including the smaller ones, the observed PL lines correspond to the superposition of the two unresolved transitions, that we may still call "A" and "B". These transitions correspond to well-width dependent combinations of $|\Gamma_{5A}\rangle$ and $|\Gamma_{5B}\rangle$ states, both dipole-allowed in σ-polarization.

Coming to the oscillator strengths –thus PL decay times– of these transitions for the narrower QWs, exchange-induced mixings now have the same kind of effect as for ZnO epilayers: reduction of oscillator strength for the ground state "A" to the benefit of "B". For each of these two transitions, using the notation in reference 3, the eigenstates of the matrix in Eq. (3) have the general form: $u(L_W)|\Gamma_{5A}\rangle + v(L_W)|\Gamma_{5B}\rangle + w(L_W)|\Gamma_{5C}\rangle$. Owing to spin-flip effects [3], the oscillator strengths in σ-polarization are proportional to $[u(L_W) + v(L_W)]^2$ multiplied by the envelope factor $f$, calculated above. The $L_W$ dependency lies essentially, here, on the variation of the α.γ off-diagonal term.

Figure 3 is a plot of a quantity inversely proportional to the calculated oscillator strengths for the two lower-lying optical transitions, versus $L_W$. The case of wide QWs (dominating QCSE) is rather simple: having nearly no exchange coupling means that the oscillator strengths of confined states built upon the $\Gamma_{5A}$ and $\Gamma_{5B}$ states are nearly equal, in σ-polarization. The only difference arises from the mixing of $\Gamma_{5B}$ and $\Gamma_{5C}$ states via spin-orbit off-diagonal terms in the Hamiltonian. Therefore, the oscillator strengths of these quantized states should follow together the well width dependence of the "envelope term". A factor has been applied to this quantity in order to fit the experimental PL decay times for the 5.2 and 7.1 nm wide QWs. For the two narrower QWs, two decay times have been plotted, resulting from the detailed measurements shown in Fig. 2.

The double-exponential PL decays measured for narrow QWs (and the single one for wide QWs) are well explained by the superposition, within the same line, of two exciton states having different recombination rates. We thus assign this observation –the first of its kind, as far as we know– to the exciton state mixing induced by the enhanced (or reduced) e-h exchange interaction.

In summary, we report the first quantitative measurement of the impact of quantum confined short range spin exchange interaction on the optical properties of wurtzitic ZnO / (Zn,Mg)O quantum wells. For well widths smaller than 3 nm, it manifests via a double exponential decay that we have explained by the redistribution of oscillator strengths between the two lower lying excitonic transitions. This effect disappears for well widths larger than 3 nm, due to the presence of the strong electric field.

**REFERENCES**

1 - K.Cho, Phys. Rev. B **14**, 4463, (1976).

2- B. Gil, et al., Jpn. J. Appl. Phys. **40** L1089,(2001).

3- B. Gil, Phys. Rev. B **64**, 201310, (2001).

4- M.Zamfirescu,et al.Phys. Rev. B **64**, 121304R (2001)

5 - M. Julier, et al. Phys. Rev. B **57**, 6791 (1998).

6- W. Ekardt, K. Lösch and D. Bimberg, Phys. Rev. B **20**, 3303 (1979).

7- G. Bastard, et al., Phys. Rev. B **26**, 1974 (1982).

8 - Y. Chen et al., Phys. Rev. B **37**, 6429 (1988).

9 - L.C. Andreani and F. Bassani, Phys. Rev. B **41**, 7536 (1990).

10 - F.Bernardini, et al. Phys. Rev. B **56**, R10024 (1997).

11 - P. Bigenwald, B.Gil and P. Boring, Physical Review B, **48**, 9122, (1993)

12 - P. Bigenwald et al., Phys. Stat. Solidi (b) **216**, 371 (1999).

13 - T. Bretagnon, et al.Physical Review B, **73**, 113304, (2006)

14 - Jin Seo Im et al., Physical Review B, **57**, R9435 (1998)

15 - C. Morhain et al., Phys. Rev. B **72** 241305R (2005).

16 - C.Morhain et al., Superlattices and Microstructures **38**, 455, (2005)

**FIGURE CAPTIONS**

**Figure 1:** Time-integrated photoluminescence spectra obtained under pulsed excitations for four different WQW widths and the band profiles, energy levels and envelope functions for electrons and heavy holes that we calculate for two extreme situations, namely narrow and wide WQWs.

**Figure 2:** Photoluminescence decay time for a wide well (right hand plot), and photoluminescence decay times for a narrow well (left hand plot). Note the different time scales in top and bottom in the left hand plot.

**Figure 3:** Measured (dots) and computed decay times versus well width. Right hand bottom insert: short-range exchange energy in $\alpha.\gamma$ meV. Left hand top insert: the computed A-B splitting versus well width to be compared with widths of photoluminescence lines.

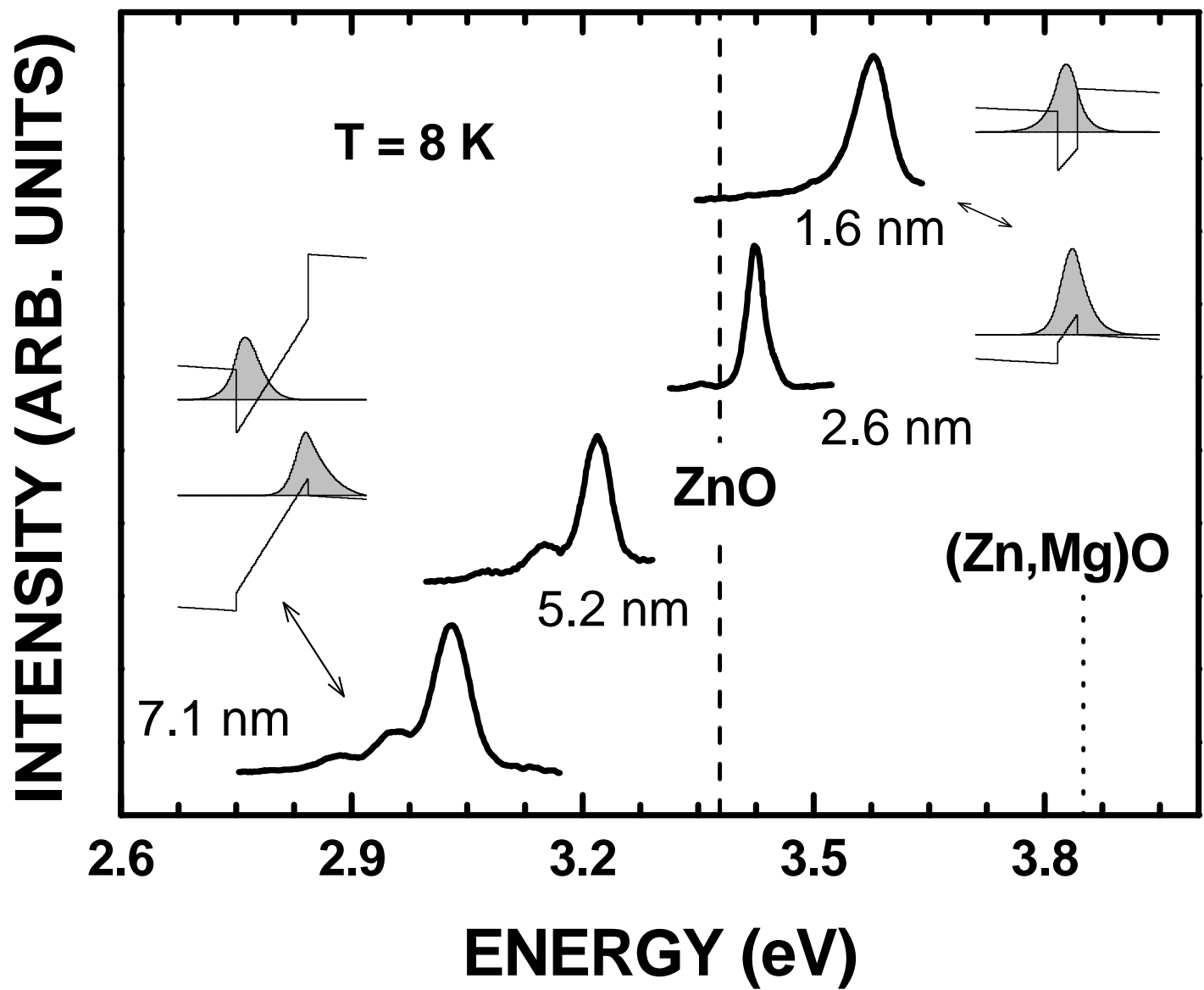

FIGURE 1

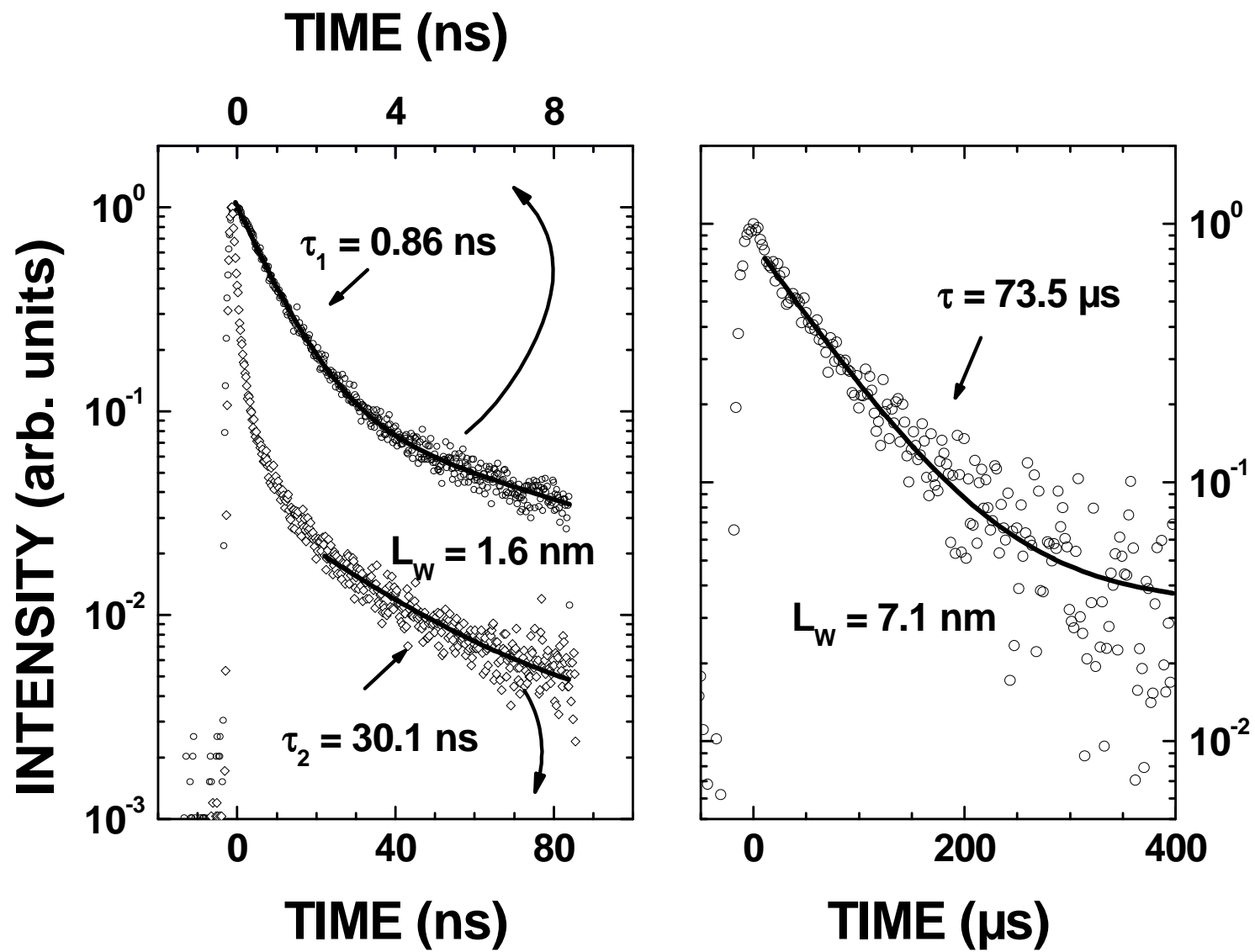

**FIGURE 2**

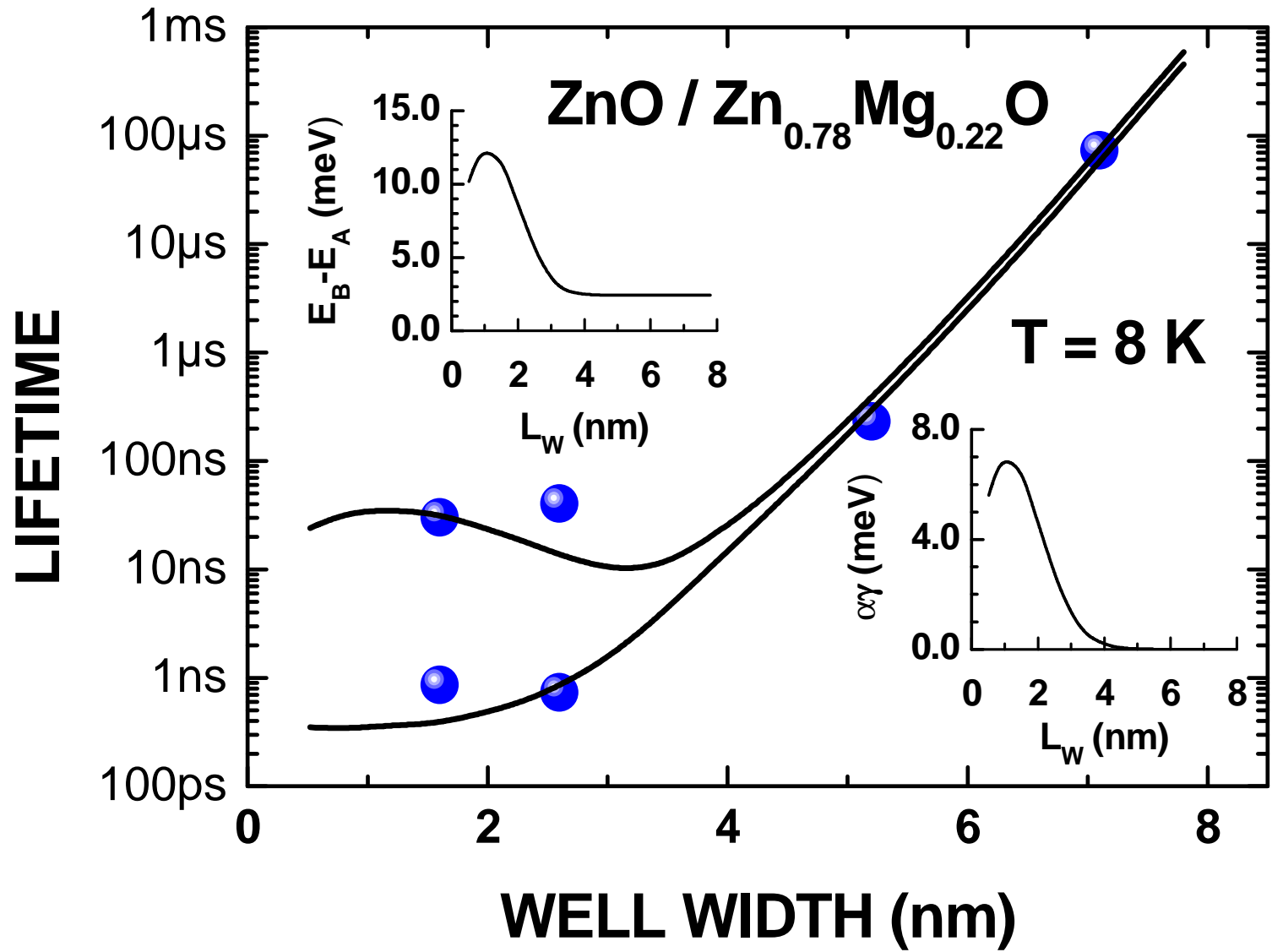

FIGURE 3